\documentclass[11pt]{JHEP3}
\usepackage{graphicx}
\usepackage{epsfig}
\usepackage{amssymb}

\newcommand{\vac}{|0\rangle}
\newcommand{\bvac}{\langle 0|}

\newcommand{\al}{\alpha}
\newcommand{\bt}{\beta}
\newcommand{\gm}{\gamma}
\newcommand{\dl}{\delta}
\newcommand{\ep}{\epsilon}
\newcommand{\zt}{\zeta}

\newcommand{\kp}{\kappa}
\newcommand{\lm}{\lambda}
\newcommand{\rh}{\rho}
\newcommand{\sg}{\sigma}

\newcommand{\ph}{\phi}
\newcommand{\vr}{\varphi}

\newcommand{\om}{\omega}

\newcommand{\half}{\frac{1}{2}}

\newcommand{\Tr}{\mbox{Tr}\,}

\newcommand{\dmu}{\partial_{\mu}}

\newcommand{\dnot}{\partial_{0}}

\newcommand{\eela}[1]{\label{#1}\end{equation}}
\newcommand{\eeala}[1]{\label{#1}\end{eqnarray}}
\newcommand{\be}{\begin{equation}}
\newcommand{\ee}{\end{equation}}
\newcommand{\bea}{\begin{eqnarray}}
\newcommand{\eea}{\end{eqnarray}}

\title{Chern-Simons number asymmetry 
from CP violation at electroweak tachyonic preheating}
\author{Jan Smit and Anders Tranberg\\
Institute for Theoretical Physics, University of Amsterdam, \\
       Valckenierstraat 65, 1018 XE Amsterdam, The Netherlands.\\
}
\preprint{ITFA-2002-51}
\abstract{We consider the creation of non-zero Chern-Simons number
in a model of the early Universe, where the Higgs field
experiences a fast quench at the end of inflation and subsequently
rolls down its potential barrier. Neglecting the expansion, we
perform numerical lattice simulations in the Abelian Higgs model
in 1+1 dimensions with an added phenomenological C and
P violating term during this stage of so-called tachyonic
preheating. The results suggest that even the sign of the Chern-Simons 
and thus baryon
number is dependent on the ratio of the Higgs to W mass. We also
discuss the appropriate choice of vacuum initial conditions for
classical simulations.}
\begin{document}

\section{Introduction}
It was proposed some time ago \cite{Garcia-Bellido:1999sv,Krauss:1999ng}
that the observed baryon asymmetry
may have been
produced in the electroweak transition at the end of an
inflationary period. The mechanism assumed resonant preheating,
in which low-momentum bands of the Higgs spectrum obtained large
occupation numbers. These would then ensure a high effective
temperature in the low-momentum gauge degrees of freedom, leading
to baryon number production via Chern-Simons number generation by
sphaleron transitions. At a later stage the system would 
thermalize at a low temperature below the electroweak scale,
leading to the well-known
suppression of the same sphaleron processes. Numerical simulations
of the 1+1 dimensional classical Abelian-Higgs model with a C and
P violating term in the action supported this idea \cite{Garcia-Bellido:1999sv}.
Recently \cite{Cornwall:2001hq}, it was proposed that electroweak production of
Chern-Simons number could even be resonant itself, with a large
boosting of the Chern-Simons number at preheating.
Another scenario that has been put forward 
exploits the production of topological defects in the Higgs field by the
Kibble mechanism, resulting in a Chern-Simons asymmetry under influence of
CP violation \cite{Krauss:1999ng,Copeland:2001qw}.

We investigate here a related mechanism that does not assume
resonant preheating or the necessity of topological defect production.
Also in this scenario inflation is assumed to end at the
electroweak scale, leaving the Universe in a cold state.
We assume that the electroweak transition was sufficiently rapid that it
resulted in a spinodal instability at essentially zero temperature.
In this paper we model it by a quench, in which
the parameter $\mu^2_{\rm eff}$ in the effective Higgs potential
$\mu^2_{\rm eff}\phi^{*}\phi+\lambda(\phi^{*}\phi)^{2}$
changes sign from positive to negative on a time scale much shorter than a
typical electroweak time.
Initially the quantum fields are in their 
semiclassical ground state at $\mu^2_{\rm eff} >0$.
Subsequently $\mu^2_{\rm eff} \to -\mu^2$ and quantum modes in the Higgs field
with momenta smaller than $\mu $ grow exponentially fast,
the spinodal instability.
After some time large occupation numbers are reached and a classical
description makes sense. One expects the growing Higgs field to generate
a growing SU(2) gauge field as well, through the classical equations of motion,
and when CP-violating interactions are operative,
a Chern-Simons asymmetry will be created. Some of this may
survive or even grow during 
the ensuing redistribution of energy over the field modes, and a 
practically frozen non-zero
Chern-Simons number may emerge if the effective
temperature is sufficiently low.

The problem we address ourselves to is: 
neglecting the expansion of the Universe
and ignoring remaining inflaton effects, how large a baryon
asymmetry is generated in the non-equilibrium process under
influence of CP violation? 
We have in mind the CP violation in the Standard
Model corresponding to the Cabibbo-Kobayashi-Maskawa matrix,
and similar terms in an extended model including neutrino mixing.
This CP violation is generally considered much too small, but the
arguments leading to this conclusion are based on dimensional
analysis involving the electroweak symmetry breaking scale
\cite{Shaposhnikov:1987pf} or
a high temperature of order of 100 GeV \cite{Rubakov:1996vz},
and the situation may be different at zero temperature and small
Higgs condensate during electroweak symmetry breaking. By assuming 
a quenching electroweak
transition at zero temperature we may expect to obtain the
largest possible baryon asymmetry in this kind of scenario.
 
In a first approach to this problem we study a 1+1 dimensional analog model
in this paper,
the abelian-Higgs model with an effective C- and P-violating interaction
of the form also used in \cite{Garcia-Bellido:1999sv} (for an early study, see \cite{Grigoriev:nv})
\be
{\cal L}_{\rm CP} = -\kp\half\epsilon_{\mu\nu}F^{\mu\nu}\phi^{*}\phi.
\label{CP1}
\ee
The more realistic
case of the SU(2)-Higgs model with effective CP-violating term 
\be
{\cal L}_{\rm CP} = -\kp \phi^{\dagger}\phi\, \Tr F_{\mu\nu} \tilde
F^{\mu\nu}
\label{CPSM}
\ee
will be reported in a future publication 
(see \cite{Smit:2002sn} for preliminary results). Real time numerical studies of the SU(2)-Higgs model in equilibrium including the above CP-violating term can be found in \cite{Ambjorn:1990pu}. An early study of the effect of a quench on the SU(2)-Higgs system is in \cite{Ambjorn:1987qu}.  
A scenario for the quench is to introduce a scalar field $\sg$,
conveniently identified with the inflaton,
with a coupling to the Higgs field given by
\be
V(\phi,\sigma)=(g_{\sg\ph}\,
\sigma^2-\mu ^{2})\phi^{*}\phi+\lambda(\phi^{*}\phi)^{2},
\ee
i.e.\ $\mu^2_{\rm eff} = g_{\sg\ph}\sg^2-\mu^2 $.
As the inflaton eventually ends its slow roll and drops to near zero,
the Higgs symmetry breaking is triggered.
This is a hybrid inflation model \cite{Linde:1991km}.
It turns out that for viability reasons
it is necessary to modify it into what is called
Inverted Hybrid Inflation \cite{Copeland:2001qw}; see also \cite{German:2001tz}
for possibilities of inflation ending at the electroweak scale.
In this study we ignore inflation and the expansion of the Universe. 

The abelian-Higgs model is introduced in section 2.
As a consequence of the spinodal instability produced by the quench,
the Higgs field acquires classical properties, 
which we review in section 3. 
The basics of this can already be found in \cite{Polarski:1995jg}.
This can be
exploited to derive realistic initial conditions for numerical
simulations using the classical approximation, as already
advocated in \cite{Smit:2001qa}. These initial conditions differ from
those e.g.\ in
\cite{Khlebnikov:1996wr}--\cite{Rajantie:2000fd}
in that they do not put power into the short-wavelength modes, 
which may lead to artificial equilibration
properties \cite{Smit:2001qa,Moore:2001zf}. We give here a detailed
derivation of these initial conditions (section 3 and 4, with a numerical
detail in the appendix). 
Related studies are in \cite{Garcia-Bellido:2001cb,Garcia-Bellido:2002aj}.
In section 5 we present our numerical results, give an interpretation
of these in terms of simple models, and analyze the possible role played
by the Kibble mechanism.
We conclude in section 6.

\section{The Abelian-Higgs model}
The Abelian-Higgs model in 1+1
dimensions is given in the continuum by the action
\bea
S&=&-\int dt\, dx
\left\{\frac{1}{4e^{2}}F_{\mu\nu}F^{\mu\nu}+
D_{\mu}\phi^{*} D^{\mu}\phi
+\frac{\mu^4}{4\lm}
- \mu^2\ph^{*}\ph + \lm(\ph^*\ph)^2
\right.\nonumber\\ && \left. \mbox{}
+\kp\half\epsilon_{\mu\nu}F^{\mu\nu}\phi^{*}\phi\right\},
\eea
where $F_{\mu\nu}= \partial_{\mu}A_{\nu}-\partial_{\nu}A_{\mu}$,
$D_{\mu}\phi=(\partial_{\mu}-iA_{\mu})\phi$,
$D_{\mu}\phi^*=(\partial_{\mu}+iA_{\mu})\phi^*$,
and $\epsilon_{01}=+1$.
We have added a P and C breaking term biasing the Chern-Simons
number:
\be
\int dx\,\kp\half\epsilon_{\mu\nu}F^{\mu\nu}\phi^{*}\phi=
2\pi\kp\dot N_{\rm CS}\phi^{*}\phi,
\ee
which is the 1+1 dimensional analog of the
CP-breaking term (\ref{CPSM}).
The Higgs and W masses are
given by $m_H^2 = 2\mu^2$, $m_W^2 = (e^2/2\lm)\, m_H^2$.

We discretize
this action on a space-time lattice in the standard way, in a periodic
spatial volume $L$. The
classical Euler-Lagrange equations of motion for $\phi$, $A_{1}$
and $A_{0}$ read in temporal gauge ($A_0=0$):
\bea
\partial_{0}'\partial_{0}\phi&=&D_{1}'D_{1}\phi
+(\mu^2 -2\lambda\phi^{*}\phi)\phi
+\kp\partial_{0}A_{1}\phi,\\
\partial_{0}'\partial_{0}A_{1}&=&
-e^{2}i(\phi^{*}D_{1}\phi-D_{1}\phi^{*}\phi)
-e^{2}
\kp\partial_{0}'(\phi^{*}\phi),
\label{dotdotA}\\
\partial_{1}'\partial_{0}A_{1}&=&
e^{2}i(\phi^{*}\partial_{0}\phi-\partial_{0}\phi^{*}\phi)
-e^{2}
\kp\partial_{1}'(\phi^{*}\phi).
\label{Gauss}
\eea
Here $\partial_{\mu}$ is the forward and $\partial'_{\mu}$ is the
backward lattice derivative,
$\dmu f(x) = [f(x+a_{\mu}\hat\mu) - f(x)]/a_{\mu}$,
$\dmu' f(x) = [f(x) - f(x-a_{\mu}\hat\mu)]/a_{\mu}$, with $a_{\mu}$ the
lattice spacing in the $\mu$-direction.
Equation (\ref{Gauss}) is the Gauss constraint. In integrated form it enforces
zero total charge in our periodic volume,
\be
\sum_x i(\phi^{*}\partial_{0}\phi-\partial_{0}\phi^{*}\phi) = 0.
\label{ztc}
\ee
The Chern-Simons number simplifies to
\be
N_{\rm CS}=-\frac{1}{2\pi}a_1 \sum_{x} A_{1}(x).
\ee
It is easy to show that the equations of motion
preserve the Gauss constraint and the total charge.

\section{Classical approximation}
\label{Class}
In the limit of small Higgs and gauge couplings
($\lambda/\mu ^2\ll 1$, $e^2/\mu ^2\ll 1$ in 1+1 dimensions), we can solve
the quantum evolution in the gaussian approximation.
Before the quench we assume the field to be
in its ground state in the potential $V(\ph)=\mu_0^2 \phi^{*}\phi$.
for simplicity we take $\mu_0 = \mu$.
The quench then just flips the sign of the potential,
$V(\phi)\rightarrow -\mu^2 \phi^{*}\phi$.
In this section we focus on the Higgs field, using the notation
appropriate for 1+1 dimensions.
The generalization to 3+1 dimensions will be obvious.
Gauge fields will be included in the next section.

Choosing periodic boundary conditions in space with volume (length)
$L$, we make a Fourier decomposition for each real field operator
$\hat\phi_{j}$, $j=1,2$, $\hat\ph = (\hat\ph_1 + i\hat\ph_2)/\sqrt{2}$, and
$\hat\pi_j = \dot{\hat\ph_j}$,
\be
\hat\phi_{j}(x)=\sum_{k}\frac{1}{\sqrt{L}}\hat\phi^{j}_{k}\,e^{ikx},
\;\;\;\;
\hat\pi_{j}(x)=\sum_{k}\frac{1}{\sqrt{L}}\hat\pi^{j}_{k}\,e^{ikx}.
\label{Fourier}
\ee
Modes with wave number
$k^{2}<\mu^2 $ will be unstable and grow exponentially,
those with $k^{2}>\mu^2 $ will not grow.
We start by expanding each real field\footnote{We drop the index $j$.}
in creation and annihilation operators
just before the quench near $t=0$,
\be
\hat\phi_{k}=\frac{1}{\sqrt{2\omega_{k}^{+}}}
(\hat a_{k} e^{-i\omega_{k}^{+}t} + \hat a_{-k}^{\dagger}e^{i\omega_{k}^{+}t}),
\;\;\;\;
\hat\pi_{k}=\frac{-i\omega_{k}^{+}}{\sqrt{2\omega_{k}^{+}}}
(\hat a_{k}e^{-i\omega_{k}^{+}t}-\hat a_{-k}^{\dagger}e^{i\omega_{k}^{+}t}).
\ee
whereas after the quench, for $t>0$, we write
\be
\hat\phi_{k}= \hat\al_{k} e^{-i\omega_{k}^{-}t}
+ \hat\bt_k e^{i\omega_{k}^{-}t}, \;\;\;\;
\hat\pi_{k}=-i\omega_{k}^{-} (\hat\al_{k}e^{-i\omega_{k}^{-}t}
- \hat\bt_k e^{i\omega_{k}^{-}t}),
\label{phpi1}
\ee
where
\be
\omega_{k}^{\pm}=\sqrt{\pm\mu^2 +k^{2}}.
\ee
Matching at time zero gives the relation
\bea
\hat\alpha_{k}&=&\frac{1}{2\sqrt{2\omega_{k}^{+}}}
\left[\left(1+\frac{\omega^{+}_{k}}{\omega_{k}^{-}}\right)\hat a_{k}
+\left(1-\frac{\omega^{+}_{k}}{\omega_{k}^{-}}\right)
\hat a_{-k}^{\dagger}\right],
\label{aldef}\\
\hat\bt_k&=&\frac{1}{2\sqrt{2\omega_{k}^{+}}}
\left[\left(1-\frac{\omega_{k}^{+}}{\omega_{k}^{-}}\right)\hat a_{k}
+\left(1+\frac{\omega_{k}^{+}}{\omega_{k}^{-}}\right)
\hat a_{-k}^{\dagger}\right].
\eea
For the stable modes $\om_k^-$ is real and $\hat\bt_k =
\hat\al_{-k}^{\dagger}$. For the unstable modes $\omega_{k}^{-}$ is
imaginary and we write $\omega^{-}_{k}=i|\omega_{k}^{-}|$ (the
opposite sign gives equivalent results). Then $\hat\al_k^{\dagger} =
\hat\al_{-k}$ and $\hat\bt_k^{\dagger} = \hat\bt_{-k}$. In both cases the
reality conditions $\hat\ph_k^{\dagger} = \hat\ph_{-k}$, $\hat\pi_k^{\dagger}
= \hat\pi_{-k}$ are satisfied.

Consider now the unstable modes, i.e.\ with $|k|$ strictly smaller
than $\mu $ and $\om_k^- = i|\om_k^-|$, which grow exponentially
fast when $|\om_k^-|t \gg 1$. Neglecting the decaying exponential
in expressions (\ref{phpi1}) gives
\be
\hat\ph_k \approx \hat\al_k e^{|\om_k^-| t}, \;\;\;\; \
\hat\pi_k \approx |\om_k^-| \hat \ph_k.
\label{phipiclas}
\ee
This strongly suggests classical behavior,
since $[\hat\al_k,\hat\al_{l}]=0$ and consequently $\hat\ph_k$ and $\hat\pi_k$
{\em commute} in this approximation. Of course, there are states
and observables for which the approximation is not valid, e.g.\
the hermitian operator $i[\hat\pi_k,\hat\ph_{-l}] = \dl_{kl}$ at all
times, being the canonical commutator.
So let us see what happens to the field correlation
functions in the physically relevant state, the initial state just
before $t=0$.

The initial state is assumed to be the ground state just
before the quench, $\vac$, which satisfies $\hat a_k\vac = 0$.
We can now find the field correlators at time $t>0$:
\bea
C_k^{\ph\ph} = \langle 0|\hat\phi_{k}\hat\phi^{\dagger}_{k}|0\rangle&=&
\frac{1}{2\omega_{k}^{+}}
\left[1+\left(\frac{\omega_{k}^{+2}}{\omega_{k}^{-2}}-1\right)
\sin^2(\om_k^- t)\right],
\label{Cphph}
\\
C_k^{\pi\pi}= \langle 0|\hat\pi_{k}\hat\pi^{\dagger}_{k}|0\rangle&=&
\frac{\omega_{k}^{- 2}}{2\omega_{k}^{+}} \left[1+
\left(\frac{\omega_{k}^{+2}}{\omega_{k}^{-2}}-1\right)
\cos^2(\om_k^- t)\right],
\\
C_k^{\ph\pi}=C_k^{\pi\ph *}
= \langle 0|\hat\ph_{k}\hat\pi^{\dagger}_{k}|0\rangle&=&
\frac{\omega_{k}^{-}}{4\omega_{k}^{+}}
\left(\frac{\omega_{k}^{+2}}{\omega_{k}^{-2}}-1\right)
\sin(2\om_k^- t) + \frac{i}{2}.
\label{Cpipi}
\eea
The correlator $\bvac \hat\pi_k^{\dagger}\hat\ph_k\vac$ follows from the
commutation relation
$\hat\ph_k\hat\pi_k^{\dagger} = \hat\pi_k^{\dagger} \hat\ph_k + i$.

We re-express these correlators in terms of
time-dependent particle numbers $n_k$, frequencies
$\om_k$, and off-diagonal particle numbers $\tilde n_k$,
as follows:
\bea
C_k^{\ph\ph}&=& (n_k + 1/2)/\om_k,\\
C_k^{\pi\pi}&=& (n_k + 1/2)\om_k,\\
C_k^{\ph\pi}&=& \tilde n_k +i/2.
\eea
These $n_k$ and $\om_k$ have proven to be robust and very useful in numerical
studies of {\em interacting} scalar fields out of equilibrium
\cite{Salle:2000hd}. Often $\tilde n_k$ is equal to zero, but here
it is not and it plays an important role in the
transition to classical behavior, as we shall see below. We note
the identity
\be
(n_k + 1/2 + \tilde n_k)(n_k +1/2 - \tilde n_k) = 1/4.
\ee

Consider again the unstable modes. Their particle numbers grow
exponentially. For $|\om_k^-| t \gg 1$ we find
\be
n_k+ \half\approx \tilde n_k \approx \half
\sqrt{\frac{\mu^4}{\mu^4-k^4}}\; e^{2\sqrt{\mu ^2-k^2}\, t},
\;\;\;\; \om_k \approx |\om_k^-| = \sqrt{\mu ^2-k^2},
\ee
and the generic field-expectation values behave as classical.
To express this more clearly, let us introduce sources $J_k^{\ph}$
and $J_k^{\pi}$ which are only nonzero for $k$ in the unstable
region, i.e.\ $k^2$ strictly smaller than $\mu^2 $:
\be
\int dx\, (J^{\ph}\hat\ph + J^{\pi} \hat\pi) =
\sum_{|k|<\mu _{\ep}}(J_{-k}^{\ph}\hat\ph_k + J_{-k}^{\pi} \hat\pi_k),
\;\;\;\;
\mu _{\ep} \equiv (1-\ep)\mu ,
\ee
where $\ep$ is a small positive number $\ll 1$. For
sufficiently large times the expectation values of products of
$\ph$' and $\pi$'s, with any operator ordering, can be calculated.
Using the approximation (\ref{phipiclas}) we find for the generating
functional
\bea
{\cal G}[J] &=& \bvac\exp\left[\int dx\,[J^{\ph}(x)\hat\ph(x)
+ J^{\pi}(x) \hat\pi(x)]\right]\vac
\nonumber\\
&\approx&
\exp\left[-\half\int dx\, dy\, J^a(x) C^{ab}_{\rm u}(x,y) J^b(y)\right],
\eea
where we summed over $a,b = \{\ph,\pi\}$,
and the dominant part of the correlator
corresponding to the unstable modes is given  by\footnote{We used
the Campbell-Baker-Haussdorf formula for
$\bvac \exp \sum_k \zt_k \hat\al_k \vac =
\exp\left(\half \sum_k\zt_{-k}\zt_k C_k^{\ph\ph}\right)$.
}
\bea
C_{\rm u}^{ab}(x,y)&=&\frac{1}{L}\sum_{|k|<\mu _\ep}e^{ik(x-y)}\,
C_{{\rm u}\, k}^{ab},
\\
C_{{\rm u}\, k} &=& \left(\begin{array}{cc}1&|\om_k^-|\\|\om_k^-|&
|\om_k^-|^2\end{array}\right) \frac{\mu^2
}{2\om_k^+\sqrt{\mu^4-k^4}}\, e^{2\sqrt{\mu^2 -k^2}\, t}.
\label{Cunst}
\eea
The matrix $C_{{\rm u}\, k}$ is singular, which reflects the fact
that $\hat\pi_k = |\om_k^-|\hat\ph_k$ in this approximation. So,
for sufficiently large times, the dominant part of the quantum
correlators can be expressed as a probability distribution in a
functional space of {\em classical} $\ph$ and $\pi$ consisting
only of unstable modes:
\bea
{\cal G}[J]&\approx& {\cal N} \int [d\ph\, d\pi]
\left[\prod_{k< \mu _{\ep}}\dl(\pi_{k} - |\om_k^-|\ph_{k})\right]
\exp\left[-\half\int
dx\,dy\, \ph^a(x) C^{-1}_{{\rm u}\, ab}(x,y)\ph^b(y)\right.
\nonumber\\ && \mbox{} + \left.
\int dx\, J^a(x)\ph^a(x)\right],
\label{pathint}
\eea
where ${\cal N}$ is such that ${\cal G}[0]=1$, and $\ph^{\ph} = \ph$,
$\ph^{\pi} = \pi$. The functional measure $[d\ph\, d\pi]$
can be explicitly (and tediously) expressed in terms of the independent
real and imaginary parts of $\pi_{k}$ and
$\ph_{k}$ with $|k| < \mu _{\ep}$,
but we will refrain from doing this here.

With the help of the classical probability distribution (\ref{pathint})
we can calculate expectation values of products of the usual observables
that have a classical correspondence, which
are represented in the quantum theory by products of
$\hat\ph$ and $\hat\pi$ (symmetrized, if needed, e.g.\ the charge density
$\hat j^0 = -\hat\pi^1\hat\ph^2 + \hat\pi^2 \hat\ph^1$.
Provided such observables are dominated by their unstable mode contribution,
we can sample the distribution (\ref{pathint}), use the
samples as initial conditions for subsequent classical dynamical evolution
into the non-linear regime, and compute expectation values by averaging
over the initial conditions.

The above prescription runs into the question: what to choose for $\ep$?
It would be nice to be able to let $\ep \to 0$, but it is clear from
(\ref{Cunst}) that we then have to do something different
near the boundary of the unstable region, because of the factor
$1/\sqrt{\mu^4-k^4}$. We propose to make the replacement
\be
C_{{\rm u}\, k} \to \mbox{Re}\, C_k =
\left(\begin{array}{cc}(n_k + \half)/\om_k& \tilde n_k\\
\tilde n_k&(n_k + \half)\om_k
\end{array}\right),
\label{ReC}
\ee
which goes over into $C_{{\rm u}\, k}$ for large $|\om_k^-| t$,
and to set $\ep = 0$. Using the real part corresponds to symmetrized
products, e.g.\
$\mbox{Re}\, C^{\ph\pi}_k =
\bvac \half(\hat\ph_k\hat\pi_k^{\dagger}
+ \hat\pi_k^{\dagger}\hat\ph_k)\vac$, which goes over
into the corresponding classical correlator in the classical limit
\cite{Aarts:1997kp}.  The delta functional in (\ref{pathint})
is to be omitted in the replacement (\ref{ReC}). It is generated
automatically in approximate form when the instability progresses.
This can be seen by introducing the variables
\be
\xi^{\pm}_{k} =
\frac{1}{\sqrt{2}}\left(\frac{\pi_{k}}{\sqrt{\om_k}}
\pm \sqrt{\om_k}\,\ph_{k} \right),
\ee
in terms of which the proposed classical distribution takes the form
\be
\exp\left[-\half\sum_{|k|<\mu }
\left(\frac{|\xi^+_k|^2}{n_k + 1/2 + \tilde n_k}
+ \frac{|\xi^-_k|^2}{n_k + 1/2 - \tilde n_k}\right)\right].
\label{xidis}
\ee
\FIGURE{\epsfig{file=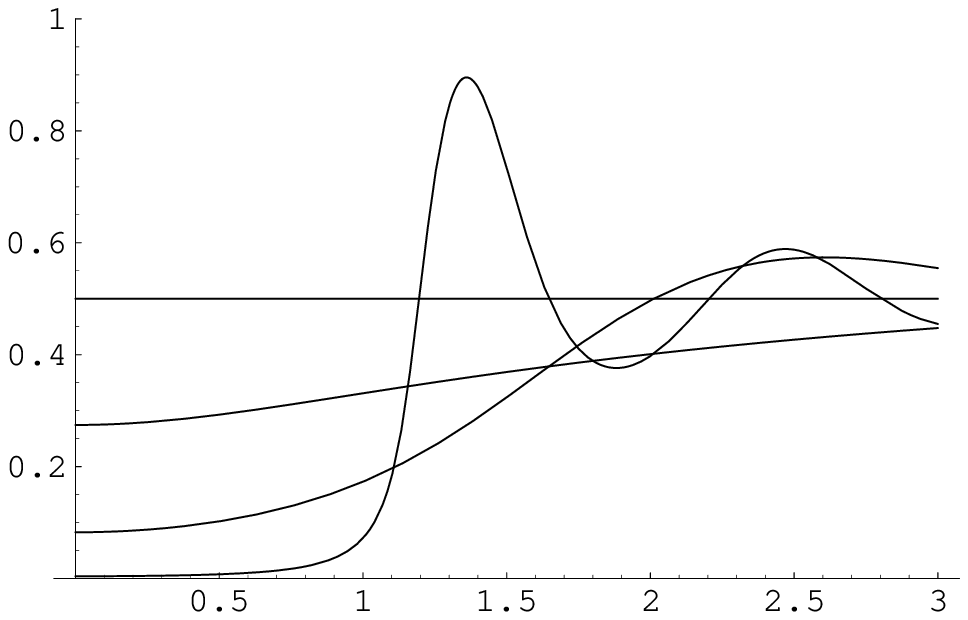,width=10cm,clip}\caption{The difference $n_k+ 1/2 -\tilde n_k$ versus $k/\mu$ at early times $t\mu = 0$, 0.3, 0.9, 2.4 (from top to bottom at $k=0$).}\label{fnmnt}}
As time progresses, $n_k+1/2-\tilde n_k\to 0$ rapidly, for
$|k|<\mu _{\ep}$, as illustrated in figure \ref{fnmnt},
and the delta functional enforcing
$\xi^-_k \propto (\pi_{k} - \om_k \ph_{k}) = 0$
in (\ref{pathint}) appears automatically.
This squeezing of the $\ph_k-\pi_k$ distribution along
$\pi_k = \om_k \ph_k$ has been studied earlier in detail in
\cite{Polarski:1995jg}, and in \cite{Garcia-Bellido:2002aj}.

We could of course drop
the $\xi^-_k$ modes altogether, since they never grow large.
However, by including them we will be able to compare with another
practical method for obtaining initial conditions in classical numerical
simulations, which will be introduced below.
Here we note that the modes with $|k| >\mu $ have $n_k + 1/2 \pm \tilde n_k$
just oscillating around 1/2, with an amplitude decaying like
$k^{-4}$. Similarly $n_k \to 1/2$ for $|k|\to \infty$. So these modes stay
in the quantum regime and we do not include them
in the initial conditions for the classical approximation.
Figure \ref{nntom} shows $n_k \pm\tilde n_k$ and $\om_k$ at time
$t=4.2\,\mu^{-1}$. 
\FIGURE{\epsfig{file=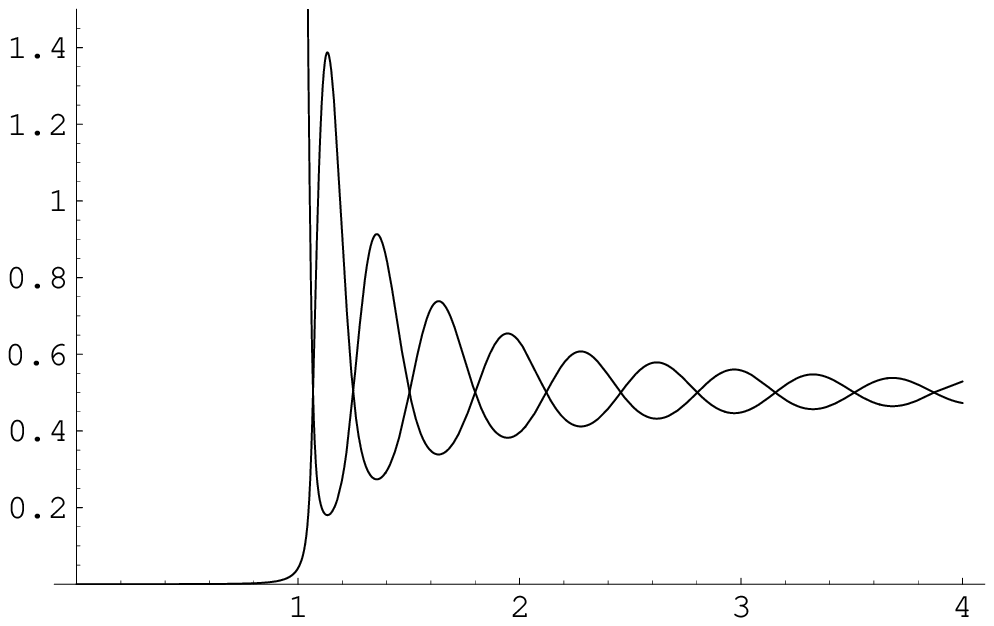,width=7cm,clip}\epsfig{file=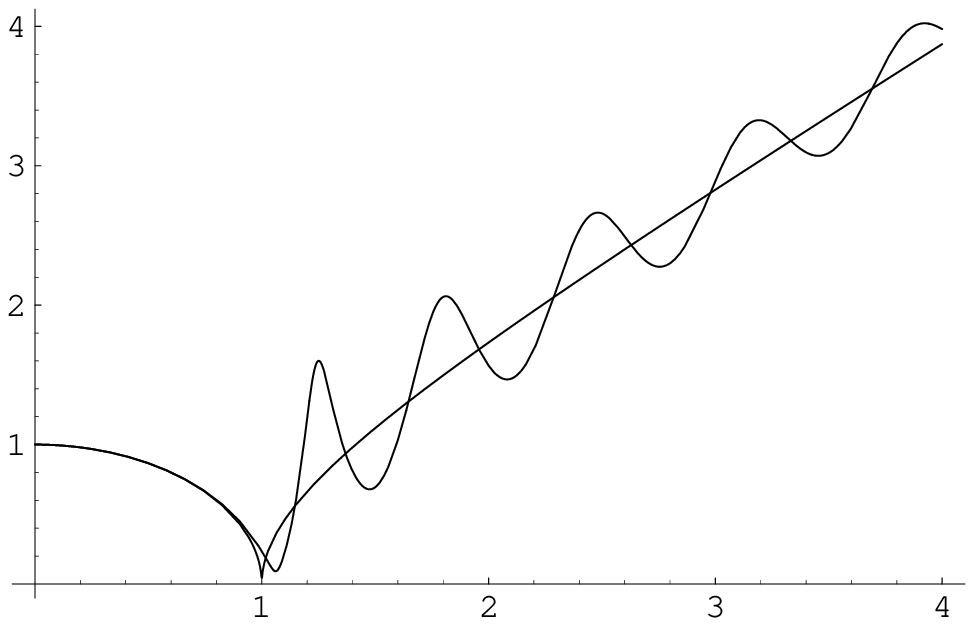,width=7cm,clip}
\caption{Left: $n_k+ 1/2 +\tilde n_k$ (coming down from
$\approx 2224$ at $k=0$) and $n_k+ 1/2 -\tilde n_k$ (practically
zero at $k=0$) versus $k/\mu$ at time $t\mu = 4.2$. Right:
$|\om_k^-|$ and $\om_k$ versus $k/\mu$ at $t\mu = 4.2$. }
\label{nntom}}

We now give an estimate of the time-span for which we have reason
to trust the free-field approximation. Consider the unstable mode
contribution to $\bvac \ph^2(x)\vac$,
$\bvac \ph^2(x)\vac_{\rm unst} \equiv \vr^2$.
We define a time $t_{\rm nl}$ (`$t$--non-linear') such that $\vr$
has grown so large that it is at the inflection point of the
potential, i.e.\ $\partial^2 V(\vr)/\partial \vr^2 = 0$, or
$\vr^2=\mu^2 /(3\lm) = v^2/3$. For times $t>t_{\rm nl}$,
non-linearities will certainly come into play. Using the
correlator (\ref{Cphph}) in the infinite volume limit this gives
the criterion, in $d$ spatial dimensions,
\be
\vr^2_d \equiv \int_{|k|<\mu } \frac{d^d k}{(2\pi)^d}\, C^{\ph\ph}_k =
\frac{\mu ^2}{3\lm}.
\label{tnl}
\ee
\FIGURE{\epsfig{file=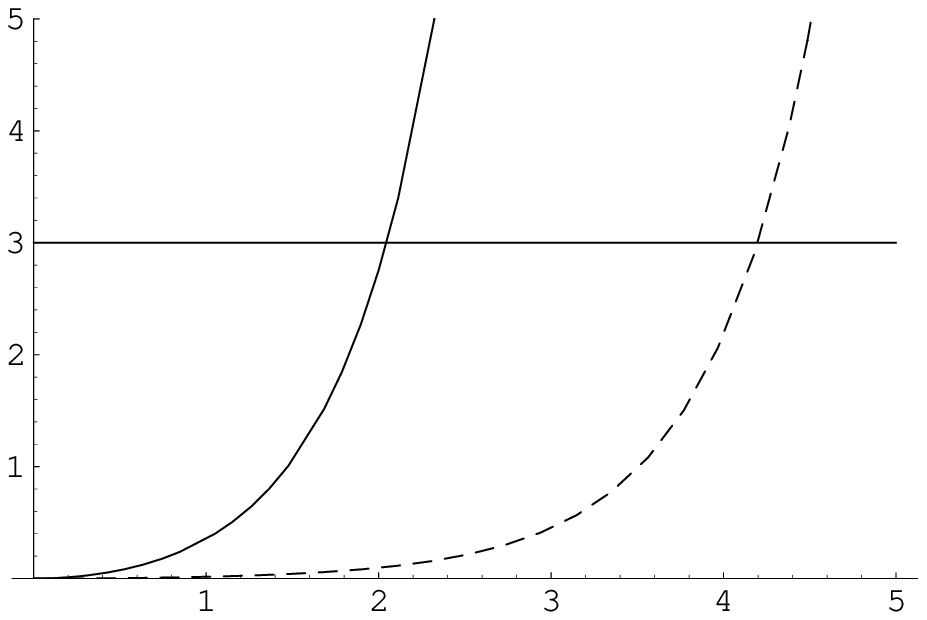,width=7cm,clip}
	\epsfig{file=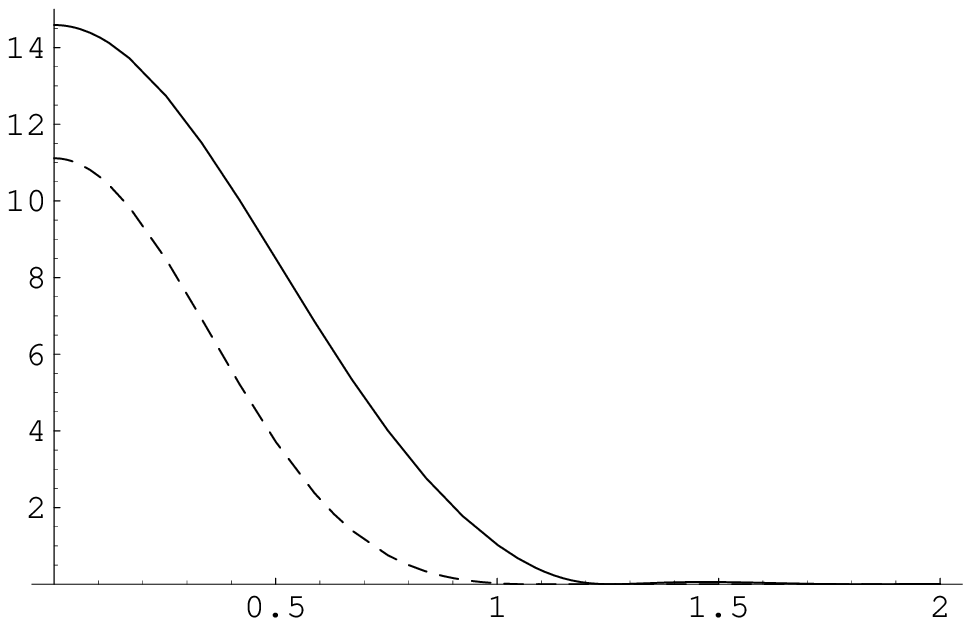,width=7cm,clip}
	\caption{Left: the function $\vr^2_d(t)/\mu^{d-1}$ versus $t\mu$ for $d=$1 (full), 3 (dashed). It reaches the value 3 at $t\mu \approx 2.05$, 4.20, respectively
for $d=1$, 3. Right: plot of $n_k^{1+1}(2.05)$ (full) and of $n_k^{3+1}(4.2)/100$ (dashed) versus $k/\mu$.}
	\label{fg1d3d}}
Figure \ref{fg1d3d} (left) shows $\vr^2_d$
as a function of time. For large time it behaves as
$\vr^2_d\approx c_d\, \mu ^{d/2-1} t^{-d/2} e^{2\mu  t}$, with
$c_1 = 0.070528$, $c_3 = 0.0056111$, which is accurate within 5\%
for $\mu t > 2$, 2.5 in one, respectively three dimensions. For
the Standard Model, $\lm=g^2 m_H^2/(8m_W^2) \approx 1/9$ for
$m_H^2/m_W^2 \approx 2$, and the corresponding value $3\approx
1/(3\lm)$ is indicated in the figure by the horizontal line. We
see that in 1+1 dimensions this line is reached earlier. Figure
\ref{fg1d3d} (right) shows the particle numbers $n_k$ at the time $t_{\rm
nl}$, for $d=1$ and 3. For small $k$ they are much larger than
one, especially in three dimensions, which justifies switching to
the classical approximation at a time somewhat before $t_{\rm nl}$.

\section{Initial conditions}
\label{Initial}
Our basic assumption is that at the end of inflation and before
the instability has set in, the system is in its time-dependent
semi-classical ground state, which we approximate by the free
vacuum of the Higgs and gauge fields corresponding to $\mu_{\rm eff}^2 > 0$.
As the unstable Higgs fields grow large, we switch to the classical
description at a roll-off time $t_{\rm ro}$ before non-linearities have
become important, $t_{\rm ro}< t_{\rm nl}$. This means that we
sample the distribution (\ref{xidis}), construct $\ph(x)$
and $\pi(x)$ from the $\xi_k^{\pm}$, and use these as initial conditions for
subsequent classical evolution. Since all Higgs field modes that
have not grown large are neglected this way (those with
$|k|>\mu$), it is natural to set the initial gauge field to zero
as well (i.e.\ $A_{1k}= \dot A_{1k} = 0$ for all $k$). However, this
would wrongly ignore the Gauss constraint.

The global Gauss constraint (\ref{ztc}) of zero total charge renders the
distribution (\ref{xidis}) non-gaussian. We take this into account by Monte Carlo methods. See the appendix for more details of this. Given a sample of the initial $\ph$ and $\pi$ with zero total charge, we then keep $A_1=0$ and satisfy Gauss' law by solving for the non-zero modes of $\dot A_1$
(its zero momentum mode is set to zero as well). Note that this brings the coupling $e^2$ into the initial conditions.

For $t>t_{\rm ro}$ the classical evolution takes place with non-zero
$e^2$ and $\lm$. But if the non-linear interactions are small at
the time $t_{\rm ro}$, then at first the correlators $C^{\ph\ph}$,
$C^{\ph\pi}$ and $C^{\pi\pi}$  are still given by the expressions
(\ref{Cphph})--(\ref{Cpipi}). So the results should not depend on
the precise value of $t_{\rm ro}$.  In fact, since the classical
evolution for $C^{ab}$ is identical to the quantum evolution in
the gaussian case, we may as well send $t_{\rm ro}$ to zero. This might
give better results in case of stronger couplings, for which the
non-linearities cannot be neglected even at small times. Since
$n_k = \tilde n_k = 0$ (and $\om_k = \om_k^+$) at $t=0$, the
denominators in the distribution (\ref{xidis}) only contain the
factor 1/2 --- which is why we call this choice of initial
conditions the `just a half' method. This is almost identical to
the choice made in \cite{Khlebnikov:1996wr}--\cite{Rajantie:2000fd},
but with the important difference
that we have a cut-off on the initialized modes
that are non-zero: $|k| < \mu$.

A completely different initialization is obtained by replacing $n_k + 1/2$
and $\tilde n_k$ in (\ref{xidis})
by a Bose-Einstein (BE) distribution,
\be
n_k +1/2 \to  \frac{1}{e^{\om_k^+/T}-1},
\;\;\;\;
\tilde n_k \to  0,
\label{nBE}
\ee
without restriction on $k$. Such an ensemble of quantal BE initial
conditions for a classical approximation may seem strange, but
they do indeed represent a free-field thermal {\em quantum} density operator
\cite{Salle:2000hd}, i.e.\ of the form
\be
\hat \rh \propto \exp\left[-\frac{1}{T}\sum_k
\left(\hat a_k^{\dagger} \hat a_k +\textstyle{\half}\right)
\om_k^+ \right],
\ee
for each real component of the Higgs field. The temperature $T$
controls the initial fluctuations that start up the tachyonic
instability. Choosing $T$ low enough, the high-momentum modes are
sufficiently suppressed to avoid Rayleigh-Jeans problems or
regularization artefacts, at least for some time after start-up,
since classical equilibration to equipartition is a slow process.

In summary, we have considered three types of initial conditions
for the classical evolution:
\begin{itemize}
\item[-] `spinodal': sampling (\ref{xidis}) at roll-off time
$2\mu^{-1} <t_{\rm ro} < t_{\rm nl}$ (assuming $t_{\rm nl} > 2\mu^{-1}$),
\item[-] `just the half': sampling (\ref{xidis}) with $n_k =
\tilde n_k = 0$ at $t=0$,
\item[-] thermal: sampling (\ref{xidis}) with the BE form
(\ref{nBE}), with $T/\mu \ll 1$.
\end{itemize}

\FIGURE{\epsfig{file=spinodal_breakdown.eps,width=10cm,clip}
	\epsfig{file=init_cond_dep.eps,width=10cm,clip}
	\caption{Top: Dependence of $\ph^{\dagger}\ph$ on the roll-off time
$t_{\rm ro}$ with `spinodal' initial conditions, including `just a
half' ($t_{\rm ro}=0$). The `thermal' case with $T/m_{H} = 0.1$ is
also shown. Bottom: $\langle N_{cs}\rangle$ in runs with three
different initial condition schemes.
	}\label{initcondplot}}
\section{Numerical results}
Most of the results presented here are for a number of lattice points $N = 512$,
lattice spacing $a m_H = a\sqrt{2}\mu = 0.3$, where $a\equiv a_1$,
temporal lattice spacing $a_0= 0.1\, a_1$,
and volume $Lm_H = 153.6$. 
We have typically used an ensemble of 1000 initial conditions and
have run for a time $t=600\, m_H^{-1}$,
keeping track of the average $\langle \ph^{\dagger}\ph\rangle$ and 
Chern-Simons number, $\langle N_{\rm CS}\rangle$.
We studied the dependence on the coupling ratio $2\lm/e^2 = m_H^2/m_W^2$
and various coupling strengths $\lm/\mu^2= 1/8$, \ldots, 1/512.
We also checked, using larger lattices ($N=10240$)  that the
final asymmetry is proportional to the physical volume, $\langle
N_{\rm CS}\rangle \propto m_H L$.

\subsection{Dependence on initial conditions}

Figure \ref{initcondplot} shows $\langle \ph^{\dagger}\ph\rangle$
and $\langle N_{\rm CS}\rangle$ at early times, for spinodal
initial conditions with various choices of the roll-off time
$t_{\rm ro}$, and a comparison is made with `just a half' and
thermal initial conditions. The $CP$-asymmetry parameter $\kp =
-0.03$. The coupling is fairly weak, $\mu^2/\lm = 8$, for which
$t_{\rm nl}\approx 1.98$ (from (\ref{tnl}) for $\vr_1^2 = 8/3$),
with $2\lm/e^2 = 1$ ($m_H=m_W$). We see the curves for $\langle
\ph^{\dagger}\ph\rangle$ already breaking away from the `just a
half' curve for somewhat smaller values of $t_{\rm ro}$ than
$t_{\rm nl}$.  We have checked that these deviations diminish for
weaker couplings. The plot for $\langle N_{\rm CS}\rangle$
contains also a comparison with the thermal method for $T/m_H =
0.1$ and $0.3$. Compared to the `just a half' curve ($t_{\rm ro} \equiv 0$),
the thermal curve has a longer `waiting time' before the initial
dip occurs, which indicates smaller fluctuations in the initial
conditions. Apparently, the effect of the vacuum fluctuations
(characterized by $n_k + 1/2 = 1/2$) is substantial in comparison
with $n_k^{\rm BE}$ at $T/m_H = 0.1$.

\subsection{Initial asymmetry}
\label{inas}
We can estimate the initial effect of the C and P breaking term
from the equations of motion. Approximating the Higgs field by its
homogeneous mode falling in the inverted quadratic potential,
$\ph(t) = \ph(0)\exp(\mu t)$, inserting this into the equation of
motion (\ref{dotdotA}) for $A_1$, with $A_1(0) = \dot A_1(0) = 0$,
neglecting the current-term
($-2e^2 |\phi|^2 A_1$) and taking into account only the $\kp$-term,
we find at $t=t_{\rm nl}$,
\be
A_1(t_{\rm nl} = - e^2 \int_{0}^{t_{\rm nl}}\phi^{*}\phi\, dt
\approx \frac{-e^2\kp}{2\mu}\, |\ph(t_{\rm nl})|^2, 
\ee
where we have neglected the seed $|\ph(0)|^2$ compared to $|\ph(t_{\rm nl})|^2$.
Using $\ph = (\ph_1 + i\ph_2)/\sqrt{2}$ and choosing the 1-axis in the
complex plane along $\ph$ (i.e.\ $\ph_2 = 0$), gives 
$|\ph(t_{\rm nl})|^2 = \vr_1^2/2 = \mu^2/6\lm$ (cf.\ (\ref{tnl})), and 
the estimate
\be
N_{\rm CS}(t_{\rm nl}) = \frac{\kp}{2\pi}\, 
\left( \frac{m_H}{m_W}\right) ^{-2}\, \frac{L m_H}{6\sqrt{2}}.
\ee
The actual value of the asymmetry in the full simulation
in the minimum of the initial dip in $N_{\rm CS}$ in figure \ref{largeres},
is indeed within a factor of two of this estimate (figure \ref{largeres}), 
and in particular, the sign is right.

\FIGURE{\epsfig{file=nores.eps,width=10cm,clip}
	\epsfig{file=largeres.eps,width=10cm,clip}
	\caption{Examples of $\langle N_{\rm CS}\rangle$ for
$m_{H}/m_{W}=0.625$ (top) and $1.0625$ (bottom). The volume $L m_H = 153.6$ and $\kp = -0.05$.}\label{largeres}}
\subsection{Dependence on $m_H/m_W$ and $\kp$}

After the initial dip, the behaviour of the Chern-Simons number depends
on the ratio of the Higgs to W mass, as shown in figure \ref{largeres}.
For $m_H/m_W$ around 0.625
$\langle N_{\rm CS}\rangle$ keeps the same sign as the initial dip
(top plot), for most others it ends up with the opposite sign (bottom plot).
The final $N_{\rm CS}$ is typically larger than the value at the bottom of the
initial dip due to what looks like resonant behavior.

We end the simulation at a time $m_{H}t=600$ when the trajectories are 
stuck (see below), and measure the final value of the average Chern-Simons 
number, $\langle N_{cs}\rangle$. The distribution of the 
1000 trajectories in such an average is just gaussian (figure \ref{probdist}). 

\FIGURE{\epsfig{file=probdist.eps,width=10cm,clip}\caption{The distribution of the Chern-Simons numbers at $tm_{H}=600$ in an ensemble of 1000 initial conditions. $\kp=-0.05$, $m_{H}/m_{W}=1.0$.}\label{probdist}}
The dependence of the final Chern-Simons number on the mass ratio
$m_H/m_W$ at fixed $\lm/\mu^2$, $m_H L$ and $\kp$ (so changing
$e^2$) is quite complicated,
see figure \ref{massdep}.
Only in a narrow range around $m_H/m_W = 0.6$
is the sign of $\langle N_{\rm CS}\rangle$
the same as that of the initial dip and the input $\kp$ (negative). For large
mass ratio the asymmetry is expected to go down because of the explicit
factor $e^2$ accompanying $\kp$ in (\ref{dotdotA}). However,
the strong dependence on the mass ratio is related to the resonant
behavior mentioned previously.

\FIGURE{\epsfig{file=massdep.eps,width=10cm,clip}\caption{Dependence of the final $\langle N_{\rm CS}\rangle$
on $m_{H}/m_{W}$.}\label{massdep}}

\FIGURE{\epsfig{file=kappadep.eps,width=10cm,clip}\caption{Dependence of the final $\langle N_{\rm CS}\rangle$ on
$\kp$ for $m_{H}/m_{W} = 1$.}\label{kappadep}}

The actual value of $\kp$ in realistic models is presumably quite small,
so it is comforting to see that the behavior of the final
$N_{\rm CS}$ as a function of $\kp$ is simply linear, see figure \ref{kappadep}.

\FIGURE{\epsfig{file=singletraj.eps,width=10cm,clip}\caption{Examples of $N_{\rm CS}$ in single trajectories. $m_{H}/m_{W}=1$.}\label{singletraj}}

\FIGURE{\epsfig{file=Ncstemp.eps,width=10cm,clip}\caption{The effective temperature calculated from the canonical
momentum of the Chern-Simons degree of freedom. }\label{Ncstemp}}

\FIGURE{\epsfig{file=manytraj.eps,width=10cm,clip}\caption{$\langle N_{\rm CS}\rangle$ for different $\kappa$ at
$m_H/m_W = 1$, $L m_H =153.6$, except for the top
curve, which represents $N_{\rm CS}/20$ in the 20 times larger volume
$L m_H = 3072$.}\label{manytraj}}

\subsection{Suppression of sphaleron wash-out}
The effective temperature at the end of tachyonic preheating should be
low enough that any Chern-Simons asymmetry created by
the tachyonic instablility does not get washed out through equilibrium-type
sphaleron processes. If the system ends up deep enough in the broken phase
the sphaleron rate will be exponentially suppressed with temperature.
We can estimate the rate by running single configurations for a long time
and monitor effective temperature and transition rate. Figure \ref{singletraj}
shows some examples.

The Chern-Simons number eventually gets almost stuck at times of order
$1000\, m_{H}^{-1}$
and afterwards the transition rate is strongly suppressed.
We estimate the temperature of the system by monitoring the average
canonical momentum of the Chern-Simons degree of freedom 
\cite{Krasnitz:1993mt,Tang:cy}
(this is the only dynamical degree of freedom in the gauge field,
as is clear e.g.\ from the formulation in the
Coulomb gauge \cite{Tang:cy}). A typical evolution of this temperature is
shown in Figure \ref{Ncstemp}. For a somewhat different value of $m_H/m_W$
it was shown in \cite{Tang:cy} that the sphaleron rate in this model is
exponentially suppressed by the sphaleron Boltzmann factor when
$\bt' \equiv \mu^3/(\lm T)$ is larger than $\simeq 7$.
In the case of figure \ref{Ncstemp} we get $\beta'\approx 10$ and so we
are deep in the broken phase with a strongly suppressed sphaleron rate.
This is also clear from estimates of the rate from pictures like
figure \ref{singletraj}.
When averaging over initial conditions, it turns out that the
average Chern-Simons number gets stuck already at $t m_{H} \approx
200$ (fig.\ \ref{manytraj}). Its distribution (fig.\ \ref{probdist}) 
then widens at a rate equal to the sphaleron rate.

Of course, in a realistic application the additional degrees of
freedom in the Standard Model and 
the expansion of the universe are expected
to lead to further suppression of the rate, such that it ends up being
negligible at practically zero temperature.

\subsection{Volume dependence}
When the volume is increased, the fluctuations 
of the final Chern-Simons number over the initial ensemble also
grow, but the fluctuations in the density $N_{\rm CS}/L$ should go down like
$L^{-1/2}$. For a very large volume, one classical realization should suffice 
to provide an accurate estimate of the density (which in the realistic case 
would determine the fermion-number density of the universe). 
We have performed simulations using a 20 times larger volume ($N = 10240$,
$L m_H = 3072$), which should result in an average density that is 
approximately the same as presented in the previous sections
(using $L m_H = 153.6$) , with a standard deviation smaller by a factor 
$\sqrt{20}$. We have checked this by using also 20 times fewer initial 
conditions ($1000 \to 50$), and found the same standard deviation for
the density as before. See the upper two curves 
and their final-time error bars in figure \ref{manytraj} (the top curve
represents the largest volume). A closer look revealed that (up to
an insignificant shift in time) the two curves are actually indistinguishable
until $t m_H \approx 13$. Also, the lattice spacing dependence is very small.

\FIGURE{\epsfig{file=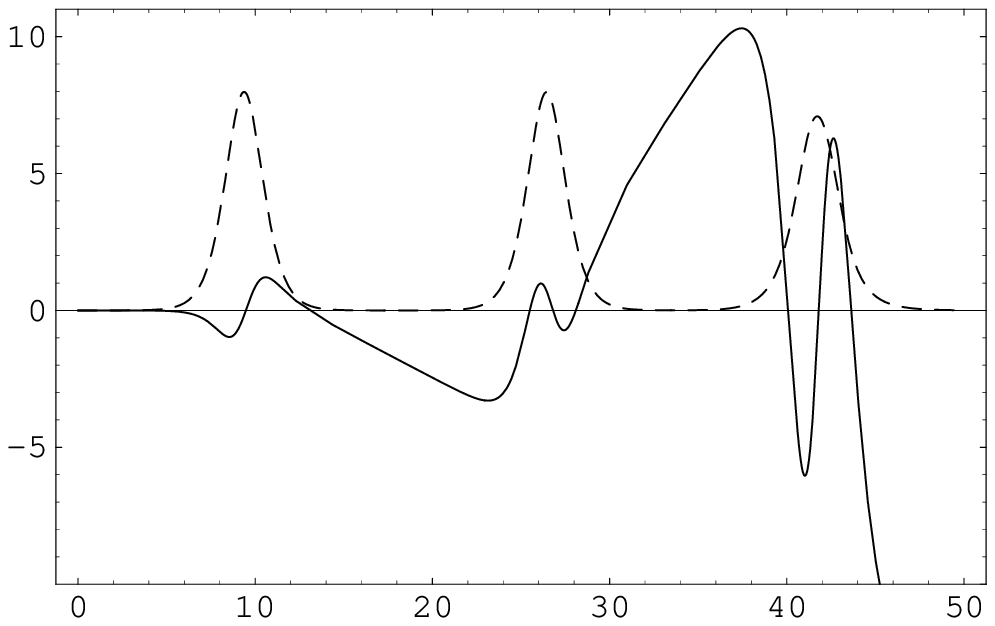,width=7cm,clip}
	\epsfig{file=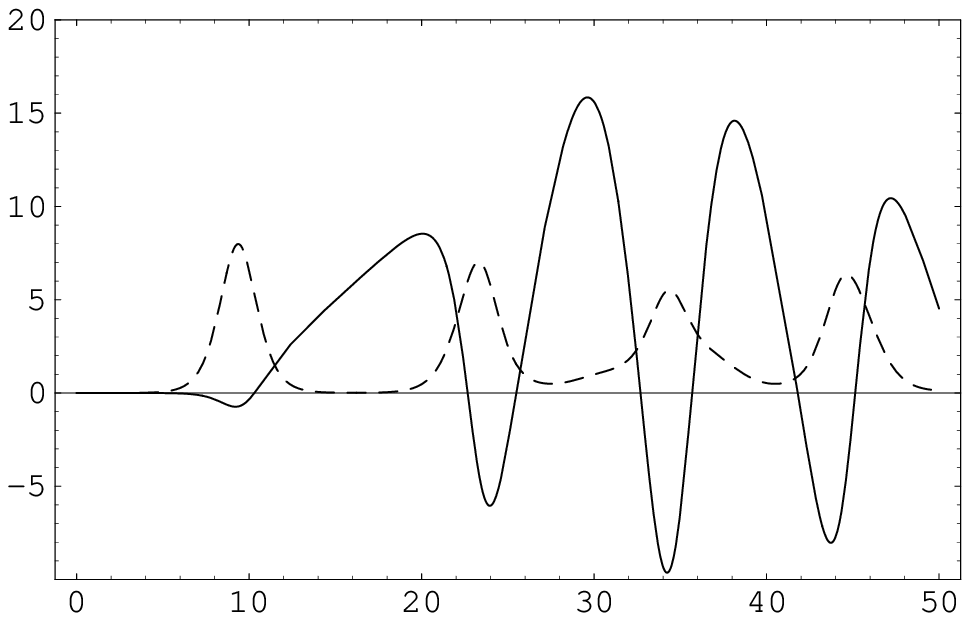,width=7cm,clip}
	\caption{Analog of figure \protect \ref{largeres} in the homogeneous
approximation; dashed: $\phi^2$ versus $t m_H$, continuous: 
$N_{\rm CS}$.
Left: $m_{H}/m_{W}=0.625$, right:  $m_H/m_W = 1.0625$.
The initial values are $\ph = 0.02$, $\dot \ph = \dot A_1 = A_1 = 0$.}
	\label{homap}}

\subsection{Modelling}
\label{mod}
In this section we attempt to interpret the data in terms of simple models
for the dynamics of $N_{\rm CS}$, or equivalently, the homogeneous gauge
field in the Coulomb gauge, $A_1$. The first model is simply the restriction
of the equations of motion to homogeneous fields, 
with the full non-linear dynamics. Without loss of generality, we can assume $\ph$ to be real.
Figure \ref{homap} shows the result for parameter values as in figure
\ref{largeres}. A striking difference with fig.\ \ref{largeres} is the 
non-linearity of the oscillations in the fields, due to the absence of
damping by the in-homogeneous modes. However, it appears that the behavior
until the second minumum of $\langle\ph^*\ph\rangle$ ($t m_H\approx 12$) 
is reasonably well 
represented. The initial dips in figure \ref{largeres} (\ref{homap}) are
-0.66 (-0.66) and -0.74 (-0.53), respectively for $m_H/m_W = 0.625$ and
1.0625. In the former case the Chern-Simons number oscillates more rapidly
when $\ph^*\ph$ is large, because the W mass is larger,
and when $\ph^*\ph$ is low again after its first
maximum, $N_{\rm CS}$ coasts along almost freely in the negative (positive)
direction in the left (right) plot of figure \ref{homap}. The homogeneous
approximation breaks clearly down already in the region of the second
minimum of $\ph^*\ph$, but 
the sign of the true final asymmetry (fig.\ \ref{largeres})
is the same as the sign of $N_{\rm CS}$ in this region
($12 \lesssim t m_H \lesssim 20$) in
the homogeneous approximation (fig.\ \ref{homap}). 

When we add damping terms $\gm \dot \ph$ and $\gm_A \dot A_1$ to the
homogeneous approximation, the resulting $\ph^2$ can be made to look
pretty much like figure \ref{largeres}, including the oscillation period
. However, the resulting $N_{\rm CS}$ then simply
performes a damped oscillation around zero. In the real simulations $N_{\rm CS}$
gets stuck in a minimum between the sphaleron energy barriers.
We have tried to model this by adding a periodic potential 
$V_{\rm s}(L A_1,\ph)$, periodic in $L A_1$ with period $2 \pi$,
and height equal to the sphaleron energy $E_{\rm s} = (2 m_H/3) v^2$,
where $v^2$ is the expectation value $2\ph^*\ph = v^2= m_H^2/2\lm$
in the classical ground state. 
The effective action then takes the form
\be
S_{\rm eff} = L \int dt\,\left[
\frac{\dot A_1^2}{2e^2} + \dot \ph^2 + \mu^2 \ph^2 - \lm \ph^4
+ \kp \dot A_1 \ph^2 + V(L A_1,\ph)\right]
\ee
Since the homogeneous $\ph$ is varying
in time we try the simplistic form $V(L A_1,\ph) \propto \ph^2$:
\be
V(L A_1,\ph) =  \ph^2 L^{-2} f(L A_1),
\ee
with the tentative conditions
\bea
f(L A_1) &=& \frac{4}{3}\, m_H L,
\;\;\;\; L A_1 = \pi,
\label{c1}
\\
&=&  (L A_1)^2 + O((L A_1)^4),
\;\;\;\; L A_1 \to 0.
\label{c2}
\eea
The first condition represents the sphaleron energy, the second takes care
of the fact that the effective $A_1$ mass is $2 e^2 \ph^2$.
A possible solution can be given in the form
$f(L A_1) = \sum_{n=0}^2 c_n(m_H L)\, \cos(n L A_1)$,
with coefficients $c_n$ depending on $m_H L$.
The equations of motion follow straightforwardly from the effective 
action, after adding also the required damping terms,
\bea
\ddot \ph + \gm \dot \ph &=& \left(m_H^2/2 - 2\lm\ph^2 
+ 2 L^{-2} f(2\pi N_{\rm CS})  + \kp \dot A_1\right)\ph,
\label{modelph}\\
\ddot N_{\rm CS} + \tilde \gm e^2 \dot N_{\rm CS} &=&
-(e^2/2\pi)\left(f'(2\pi N_{\rm CS}) \ph^2
+ 2\kp L \ph\dot\ph\right)
\label{modelN}
\eea
where we assumed the damping for $A_1$ to be proportional to $e^2$.

However, for $m_H L$ of order 1 (the sphaleron size) and $\kp = 0.05$, 
the $\kp$-term is much too small to push the Chern-Simons number
over the sphaleron barrier and the resulting asymmetry is zero. For 
larger volumes the barrier itself is too high to get $N_{\rm CS}$ sufficiently
away from zero; the $f'$-term in (\ref{modelN})
becomes very large near the top of the barrier
(cf.\ (\ref{c1})).
The $f$-term in (\ref{modelph}) may actually be neglected for large $L$.

To proceed, we shall interpret the above equations as
effective equations for the Chern-Simons {\em density},
with an effective potential $f$ that is not constrained by the barrier condition
(\ref{c1}), and that is even time-dependent. We have to incorporate the
fact that at early times $t \lesssim 13\, m_H^{-1}$ the quadratic form
$L^{-2} f(LA_1) = A_1^2$ gives a reasonable description of the data. 
The bottom of the dips in figure \ref{homap} is deeper than the sphaleron
value $-1/2$. This means that the sphalerons have not appeared yet, because
$N_{\rm CS}$ bounces back (on the $A_1^2$ potential)
instead of rolling down the hill on the other side of a (lower) sphaleron
barrier. We model this by an effective potential that changes from quadratic
to periodic, e.g.\
\bea
L^{-2} f(L A_1) &\to& \ell^{-2}[2-2\cos(\ell A_1)],\label{sphpot1}
\\
\ell(t) &=& \ell_0 + \ell_1[1+\tanh(\dl(t-t_0))]/2,\label{sphpot2}
\eea
with parameters 
such that $\ell(t)$ is very small at $t=0$ (and consequently 
$L^{-2} f(LA_1) \approx A_1^2$) and reasonably large at $t=\infty$,
such that $A_1$ is able to hop over the barrier. For simplicity we set
$\ell_1 = 153.6\, m_H^{-1}$, the value of $L$ in most figures. 
Then, with $\ell_0 = 0.01\, m_H^{-1}$, $t_0 = 16\, m_H^{-1}$,
$\dl = 0.16\, m_H$, and damping
coefficients $\gm = 0.14\, m_H$, $\tilde \gm = 0.8\, m_H^{-1}$, we get the
result shown in figure \ref{homap2}. 
The other parameters are as in figure \ref{largeres} and $\ref{homap}$.

\FIGURE{\epsfig{file=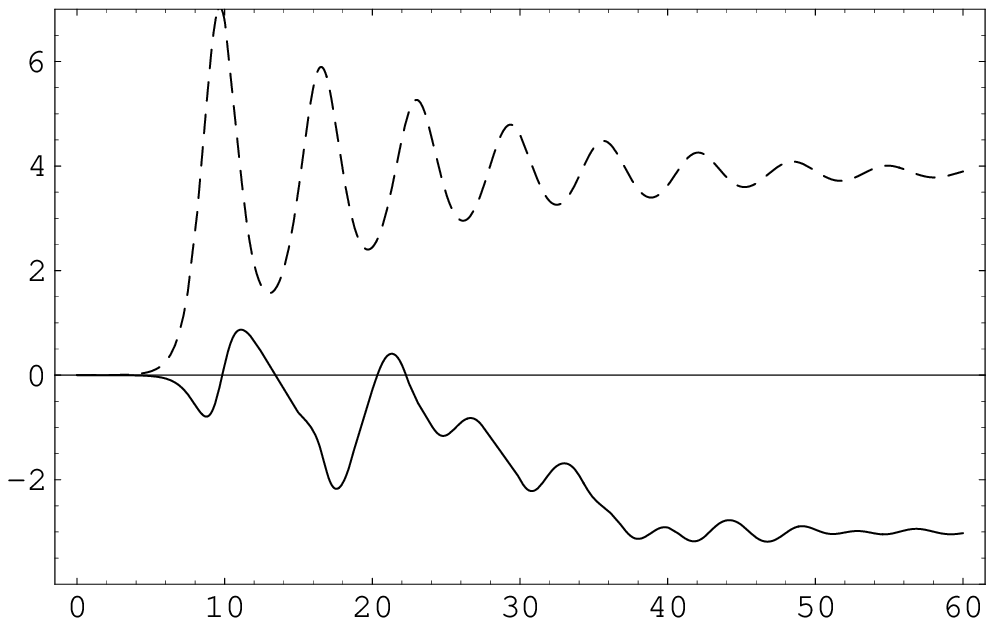,width=7cm,clip}
	\epsfig{file=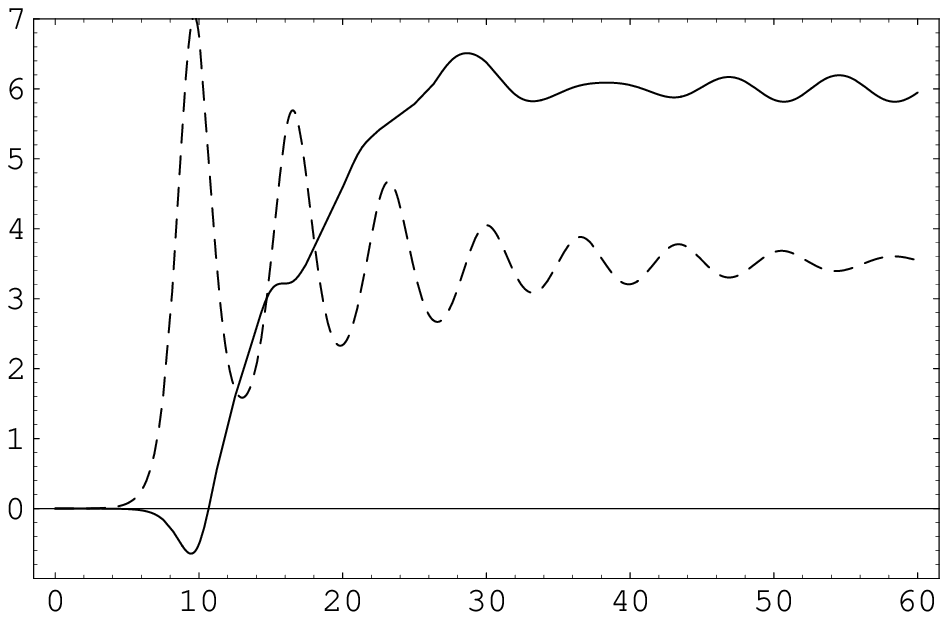,width=7cm,clip}
	\caption{Analog of figure \protect \ref{homap} 
including the time-dependent sphaleron-like potential (\ref{sphpot1},\ref{sphpot2}); 
dashed: $\phi^2$ versus $t m_H$, continuous: 
$N_{\rm CS}$.
Left: $m_{H}/m_{W}=0.625$, right:  $m_H/m_W = 1.0625$.}
	\label{homap2}}

Although figure \ref{homap2} appears to capture the qualitative behavior of
the process, we should keep in mind that the model system is quite chaotic,
and a small change of e.g.\ damping coefficients can change the result. 
This could be avoided by considering an ensemble of initial conditions.
The real simulation lacks of course the arbitrary parameters of the modelling,
and it is also more subtle, e.g.\ in that
the resonance-like behavior seen in figure \ref{largeres}
is not very well captured by the modelling.

\subsection{Role of topological defects}
The mechanism proposed in \cite{Krauss:1999ng,Copeland:2001qw} assumed that the electroweak
transition produced topological defects in the Higgs field, by the 
Kibble mechanism, with winding numbers averaging to zero in absence of 
CP violation. With CP-violating there would then subsequently be a bias
towards non-zero average winding number, with a corresponding Chern-Simons
number characterizing the vacuum. Translated to our current notation the
asymmetry would then be given by a formula of the form 
$n_{\rm CS} \equiv N_{\rm CS}/L = c\, \kp\, n_{\rm d}$, where $n_{\rm d}$ is the
density of defects and $c$ a dimensionless factor.

However, although we do not doubt the fact that there will be a non-zero
density of topological defects shortly after the transition, this does not play
a role in our interpretation of the asymmetry as presented in the previous
section. With the effective C and P violation used here, the all important
initial asymmetry 
(section \ref{inas}) is already about 10\% or more of the final value,
and it has nothing to do with the winding-number density of the Higgs field;
it depends only on $\kp\dnot (\ph^*\ph)$.

\FIGURE{\epsfig{file=Kibblenoreal.eps,width=7cm,clip}
	\epsfig{file=Kibblereal.eps,width=7cm,clip}
	\caption{Winding number of the Higgs field $N_w$, Chern-Simons number 
$N_{\rm CS}$
and $\ph^*\ph$ in two large-volume ($L mH = 3072$)
simulations, one with the usual
thermal initial conditions (left), 
and one for which the initial Higgs winding
is suppressed (right).}\label{check}}

As a check
we performed a simulation with initial conditions such, that there is no 
Kibble mechanism. They were obtained from the thermal initial values by
rotating $\ph(x)$ and $\pi(x)$ to the positive real axis, 
(i.e.\ $(\mbox{Re}\, \ph, \mbox{Im}\,\ph)) \to (|\ph|,0)$, 
$(\mbox{Re}\, \pi, \mbox{Im}\,\pi)) \to (|\pi|,0)$), and subsequently
enforcing Gauss's law. This has the effect that initially there is no
topological winding number in the Higgs field.
Figure \ref{check} shows an example of the effect
on one large-volume configuration, for the $m_H=m_W$ case.
We have also plotted the winding number of the Higgs field\footnote{Using the
gauge-invariant winding number of  
\cite{Kajantie:1998bg} gave identical results}.
\be
N_w=\frac{1}{2\pi}\sum_{x}\left[\alpha(x+a_{1})-\alpha(x)\right]_{-\pi}^{\pi}
\ee
with $\ph(x)=\rho(x)e^{i\alpha(x)}$, and where $\left[\ldots\right]_{-\pi}^{\pi}$ denotes taking the value modulo $2\pi$.
The artificially non-winding initial configurations are evidently much less
random, and we see indeed that $\ph^*\ph$  in the right-hand plot of fig.\
\ref{check} suffers less damping in its first half-period
than in the left-hand plot (which resembles
closely the corresponding plot in fig.\ \ref{largeres}). Note also the deeper
and broader
minimum around $t m_H = 13$, which reminds us of that in figure
\ref{homap}. However, we note in particular the complete absence of 
Higgs-winding in the right-hand plot of fig.\ \ref{check} in $0 < t m_H < 12$,
as compared to the somehat noisy winding in this time-interval 
in the left-hand plot. After time $t m_H \simeq 30$, it appears that the
Higgs-winding number is pulled along by the Chern-Simons number (bringing down
the magnitude of the
covariant derivative $D_1 \ph$), such that in both plots $N_{\rm CS}
\approx N_w$. Despite the large initial deviations, the final asymmetry
is semi-quantitatively unchanged with the 
artificially winding-suppressing initial conditions.

We conclude that 
it is the gauge field that,
after being biased into the `initial dip', pulls the Higgs phase along,
and the picture of a multitude of
topological defects un-winding under a biasing
C(P) asymmetry in the equations of motion is not relevant for the final
asymmetry in the present model.

\section{Conclusion}

The 1+1 D abelian-Higgs model illustrates nicely that a sizable Chern-Simons
asymmetry can be produced by a tachyonic electroweak transition
under the influence of the usual effective C- and P-violating interaction.
The numerical results can be summarized as
\be
n_{\rm CS} = \frac{\langle N_{\rm CS}\rangle}{L} = -0.7\, \kp\, m_H,
\ee
at $m_H/m_W=1$ with the dependence on Higgs to W mass ratio as shown in 
figure \ref{massdep}. 
The linearity in $\kp$ enabled us to carry out the simulations at a 
much larger value than might be expected in realistic applications.

Besides initial conditions that are motivated by
the quantum-to-classical transition in the gaussian approximation,
we also used low-temperature thermal noise for generating initial 
configurations. We found that the quantitative results do depend 
moderately on the choice of initial conditions, but not
the qualitative outcome.

It came as a surprise to us that
(even the sign of) the final Chern-Simons number and thus the corresponding 
baryon number 
asymmetry is very sensitive to the Higgs mass. An interpretation of this
intriguing effect was given in section \ref{mod}.

The mechanism for the generation of the asymmetry here is different from
that suggested in \cite{Garcia-Bellido:1999sv,Krauss:1999ng,Copeland:2001qw}, 
since neither resonant preheating nor Kibble-like generation of topological
defects plays a crucial role in the model studied here.

We have neglected here any effects related to the dynamics of
the expansion of the universe, because our primary aim is to see
the order of magnitude of the asymmetry that can be generated, given a form
of CP violation. It will be very interesting to see similar results
in the physically relevant SU(2)-Higgs theory in 
3+1 dimension \cite{Smit:2002sn}.

\appendix

\section{Implementing zero total charge}
As discussed in section \ref{Initial},
we need to generate the distribution (\ref{xidis}) subject to the constraint
of zero total charge, $\sum_x e^2 (\ph^*\pi-\pi^*\ph)$, or in terms of
Fourier variables (defined as in (\ref{Fourier})),
\be
Q \equiv
\sum_{k}\left(\pi_{k}^{1^*}\phi_{k}^{2}-\pi_{k}^{2*}\phi_{k}^{1}\right)=0.
\label{zc}
\ee
On the spatial lattice with an even number of $N$ sites, $x=ma$,
$m=0$, \ldots, $N-1$, the wave vectors can be chosen to take the
values
\be
k=\frac{2\pi n}{Na}, \;\;\;\;\;\; n= -N/2+1,...,N/2.
\ee
The reality of the fields $\ph^j(x)$ and $\pi^j(x)$, $j=1,2$, and
the fact that $\exp(ikx)$ is real for $n=0$, $N/2$, imply that we
can write
\bea
\phi_{k}^{j}&=&\ph_{-k}^{j*} =\frac{1}{\sqrt{2\omega_{k}}}
\left(a_{k}^{j}+ib_{k}^{j}\right),
\;\;\;\;
n=1, \ldots, \frac{N}{2}-1,
\\
\pi_{k}^{j}&=&\pi_{-k}^{j*} =\sqrt{\frac{\omega_{k}}{2}}
\left(c_{k}^{j}+id_{k}^{j}\right),
\;\;\; n=1, \ldots, \frac{N}{2}-1,
\\
\phi_{k}^{j}&=&\frac{1}{\sqrt{\omega_{k}}}\, a^j_{k},
\;\;\;\;
\pi_{k}^{j}=\sqrt{\omega_{k}}\, c^j_{k},
\;\;\;\;
n=0,\; \frac{N}{2},
\eea
where the real $a$'s, \ldots, $c$'s, are independent variables. In
terms of these the zero-charge condition (\ref{zc}) takes the form
\be
Q = c^{1}_{0}a^{2}_{0}-c^{2}_{0}a^{1}_{0}
+c^{1}_{N/2}a^{2}_{N/2}-c^{2}_{N/2}a^{1}_{N/2}
+\sum_{k=1}^{N/2-1} \left( c^{1}_{k}a^{2}_{k}-c^{2}_{k}a^{1}_{k}
+d^{1}_{k}b^{2}_{k}-d^{2}_{k}b^{1}_{k}
\right)=0.
\ee
\FIGURE{\epsfig{file=initdist.eps,width=10cm,clip}
	\caption{The distribution of (plain) $\phi_{0}^{1}$ (the constrained variable) and
of (checkered) $\phi_{0}^{2}$ for a set of 500 initial conditions for a
thermal ensemble at temperature $T/m_{H}=0.1$. Overlaid, a
gaussian with the relevant width.}\label{MCdist}}
To implement the zero charge constraint, the probability
distribution (\ref{xidis}) has to be multiplied by
$\dl(Q(a,b,c,d))$, which means that it no longer depends
quadratically on the $a_k^j$,\ldots, $c_k^j$. For example, for
thermal initial conditions $\tilde n_k = 0$, and we have
\bea
P(a,b,c,d) &\propto&
\delta(Q) \exp\left[
-\frac{1}{2}\sum_{k=1}^{N/2-1}\sum_{j=1}^2
\left(\frac{a^{j2}_{k}+b^{j2}_{k}+c^{j2}_{k}+d^{j2}_{k}}{n_{k}+1/2}\right)
\right.\nonumber\\&&\left.\mbox{}
-\frac{1}{2}\sum_{j=1}^2\left(\frac{a^{j2}_{0}+c^{j2}_{0}}{n_{0}+1/2}
+\frac{a^{j2}_{N/2}+c^{j2}_{N/2}}{n_{N/2}+1/2}\right)
\right]
\eea
We can now integrate out one variable, say $a^{1}_{0}$ to get rid
of the $\delta$ function
\bea
P &\to&
\frac{1}{|c^{2}_{0}|}\exp\left[
-\frac{1}{2}\sum_{k,j}
\left(\frac{a^{j2}_{k}+b^{j2}_{k}+c^{j2}_{k}+d^{j2}_{k}}{n_{k}+1/2}\right)
\right.\nonumber\\&&\left.\mbox{}
-\frac{1}{2}\sum_{j}\left(\frac{c^{j2}_{0}}{n_{0}+1/2}
+\frac{a^{j2}_{N/2}+c^{j2}_{N/2}}{n_{N/2}+1/2}\right)-\frac{1}{2}
\frac{(\tilde Q/c^{2}_{0})^{2}}{n_{0}+1/2}\right],
\label{finaldist}
\eea
where
\be
\tilde{Q}(a,b,c,d) = Q(a,b,c,d) +c^{2}_{0}a^{1}_{0}.
\ee
This distribution is no longer gaussian in the remaining $a_k^j$,
\ldots, $c_k^j$. We sample it using Monte-Carlo methods.
Sampling a distribution with a $\delta$-function is notoriously
difficult. However if we have enough variables on which to
``distribute'' the constraint, the deviation from a gaussian
distribution is expected to be small, which is indeed the case,
see figure \ref{MCdist}. Our choice of dependent variable
($a_0^1$) should not matter, in principle, but only with an ideal
Monte-Carlo algorithm. As it turned out, the number of variables
was sufficiently large that no problem was encountered.

Having produced a realization, we solve for $a^{1}_{0}$ and so
construct a field configuration $\{\ph^j(x),\pi^j(x)\}$ with zero
$Q$. As mentioned in section \ref{Initial}, the gauge field
configuration then follows from $A_1(x) = 0$ and determining
$\dnot A_1(x)$ by solving the Gauss constraint.

\acknowledgments
We like to thank Mischa Sall\'e and Jeroen Vink for useful discussions.
This work was supported in part by FOM. AT enjoyed support from the ESF network COSLAB.


\begin{thebibliography}{6}

\bibitem{Garcia-Bellido:1999sv}
J.~Garc\'{\i}a-Bellido, D.~Y.~Grigoriev, A.~Kusenko and M.~E.~Shaposhnikov,
Phys.\ Rev.\ D {\bf 60} (1999) 123504
[arXiv:hep-ph/9902449].

\bibitem{Krauss:1999ng}
L.~M.~Krauss and M.~Trodden,
Phys.\ Rev.\ Lett.\  {\bf 83} (1999) 1502
[arXiv:hep-ph/9902420].

\bibitem{Cornwall:2001hq}
J.~M.~Cornwall, D.~Grigoriev and A.~Kusenko,
Phys.\ Rev.\ D {\bf 64} (2001) 123518
[arXiv:hep-ph/0106127].

\bibitem{Copeland:2001qw}
E.~J.~Copeland, D.~Lyth, A.~Rajantie and M.~Trodden,
Phys.\ Rev.\ D {\bf 64} (2001) 043506
[arXiv:hep-ph/0103231].

\bibitem{Shaposhnikov:1987pf}
M.~E.~Shaposhnikov,
Nucl.\ Phys.\ B {\bf 299} (1988) 797.

\bibitem{Rubakov:1996vz}
V.~A.~Rubakov and M.~E.~Shaposhnikov,
Usp.\ Fiz.\ Nauk {\bf 166} (1996) 493
[Phys.\ Usp.\  {\bf 39} (1996) 461]
[arXiv:hep-ph/9603208].

\bibitem{Grigoriev:nv}
D.~Y.~Grigoriev, M.~E.~Shaposhnikov and N.~Turok,
Phys.\ Lett.\ B {\bf 275} (1992) 395.

\bibitem{Ambjorn:1990pu}
J.~Ambjorn, T.~Askgaard, H.~Porter and M.~E.~Shaposhnikov,
Nucl.\ Phys.\ B {\bf 353} (1991) 346.

\bibitem{Ambjorn:1987qu}
J.~Ambjorn, M.~Laursen and M.~E.~Shaposhnikov,
Phys.\ Lett.\ B {\bf 197} (1987) 49.

\bibitem{Linde:1991km}
A.~D.~Linde,
Phys.\ Lett.\ B {\bf 259} (1991) 38.


\bibitem{German:2001tz}
G.~German, G.~Ross and S.~Sarkar,
Nucl.\ Phys.\ B {\bf 608} (2001) 423
[arXiv:hep-ph/0103243].

\bibitem{Smit:2001qa}
J.~Smit, J.~C.~Vink and M.~Salle,
arXiv:hep-ph/0112057.
M.~Salle, J.~Smit and J.~C.~Vink,
Nucl.\ Phys.\ Proc.\ Suppl.\  {\bf 106} (2002) 540
[arXiv:hep-lat/0110093].

\bibitem{Khlebnikov:1996wr}
S.~Y.~Khlebnikov and I.~I.~Tkachev,
Phys.\ Lett.\ B {\bf 390} (1997) 80.

\bibitem{Khlebnikov:1996zt}
S.~Y.~Khlebnikov and I.~I.~Tkachev,
Phys.\ Rev.\ Lett.\  {\bf 79} (1997) 1607.

\bibitem{Khlebnikov:1996mc}
S.~Y.~Khlebnikov and I.~I.~Tkachev,
Phys.\ Rev.\ Lett.\  {\bf 77} (1996) 219.

\bibitem{Felder:2001kt}
G.~N.~Felder, L.~Kofman and A.~D.~Linde,
Phys.\ Rev.\ D {\bf 64} (2001) 123517.

\bibitem{Felder:2000hj}
G.~N.~Felder, J.~Garcia-Bellido, P.~B.~Greene, L.~Kofman, A.~D.~Linde and I.~Tkachev,
Phys.\ Rev.\ Lett.\  {\bf 87} (2001) 011601.

\bibitem{Felder:2000hr}
G.~N.~Felder and L.~Kofman,
Phys.\ Rev.\ D {\bf 63} (2001) 103503.

\bibitem{Rajantie:2000nj}
A.~Rajantie, P.~M.~Saffin and E.~J.~Copeland,
Phys.\ Rev.\ D {\bf 63} (2001) 123512.

\bibitem{Rajantie:2000fd}
A.~Rajantie and E.~J.~Copeland,
Phys.\ Rev.\ Lett.\  {\bf 85} (2000) 916.

\bibitem{Moore:2001zf}
G.~D.~Moore,
JHEP {\bf 0111} (2001) 021
[arXiv:hep-ph/0109206].

\bibitem{Polarski:1995jg}
D.~Polarski and A.~A.~Starobinsky,
Class.\ Quant.\ Grav.\  {\bf 13} (1996) 377
[arXiv:gr-qc/9504030].

\bibitem{Garcia-Bellido:2001cb}
J.~Garcia-Bellido and E.~Ruiz Morales,
Phys.\ Lett.\ B {\bf 536} (2002) 193
[arXiv:hep-ph/0109230].

\bibitem{Garcia-Bellido:2002aj}
J.~Garc\'{\i}a-Bellido, M.~Garc\'{\i}a P\'erez and A.~Gonz\'alez-Arroyo,
arXiv:hep-ph/0208228.

\bibitem{Smit:2002sn}
J.~Smit and A.~Tranberg,
arXiv:hep-ph/0210348.

\bibitem{Salle:2000hd}
M.~Salle, J.~Smit and J.~C.~Vink,
Phys.\ Rev.\ D {\bf 64} (2001) 025016
[arXiv:hep-ph/0012346].


\bibitem{Aarts:1997kp}
G.~Aarts and J.~Smit,
Nucl.\ Phys.\ B {\bf 511} (1998) 451
[arXiv:hep-ph/9707342].

\bibitem{Krasnitz:1993mt}
A.~Krasnitz and R.~Potting,
Phys.\ Lett.\ B {\bf 318} (1993) 492
[arXiv:hep-ph/9308307].

\bibitem{Tang:cy}
W.~H.~Tang and J.~Smit,
Nucl.\ Phys.\ B {\bf 540} (1999) 437
[arXiv:hep-lat/9805001].

\bibitem{Kajantie:1998bg}
K.~Kajantie, M.~Karjalainen, M.~Laine, J.~Peisa and A.~Rajantie,
Phys.\ Lett.\ B {\bf 428} (1998) 334
[arXiv:hep-ph/9803367].
\end{thebibliography}
\end{document}